\documentclass{article}

\usepackage{PRIMEarxiv}

\usepackage[utf8]{inputenc} % allow utf-8 input
\usepackage[T1]{fontenc}    % use 8-bit T1 fonts
\usepackage{hyperref}       % hyperlinks
\usepackage{url}            % simple URL typesetting
\usepackage{booktabs}       % professional-quality tables
\usepackage{amsfonts}       % blackboard math symbols
\usepackage{nicefrac}       % compact symbols for 1/2, etc.
\usepackage{microtype}      % microtypography
\usepackage{lipsum}
\usepackage{fancyhdr}       % header
\usepackage{graphicx}       % graphics
\graphicspath{{media/}}     % organize your images and other figures under media/ folder
\usepackage{longtable}
\usepackage{placeins}

\title{MindOpt Adapter for CPLEX\\ Benchmarking Performance Analysis
}
\author{
  Mou Sun \qquad Tao Li \qquad Wotao Yin \\[5pt]
  Decision Intelligence Lab \\
  DAMO Academy\\
  Alibaba Group\\
  \texttt{\{mou.sunm, coy.lt, wotao.yin\}@alibaba-inc.com} \\
  \texttt{ } \\
  \today \\
}

\begin{document}

\maketitle

\begin{abstract}
This report provides a comprehensive analysis of the performance of MindOpt Adapter for CPLEX 12.9 in benchmark testing. CPLEX, recognized as a robust Mixed Integer Programming (MIP) solver, has faced some scrutiny regarding its performance on MIPLIB 2017 when configured to default settings. MindOpt Adapter aims to enhance CPLEX's performance by automatically applying improved configurations for solving optimization problems. Our testing demonstrates that MindOpt Adapter for CPLEX yields successfully solved 232 of the 240 problems in the MIPLIB 2017 benchmark set. This performance surpasses all the other solvers in terms of the number of problems solved and the geometric mean of running times. The report provides a comparison of the benchmark results against the outcomes achieved by CPLEX under its default configuration.
\end{abstract}

\section{Introduction}

Mixed Integer Linear Programming (MILP) is a type of mathematical optimization problem characterized by linear constraints, a linear objective function, a mixed set of continuous and integer variables. They commonly arise in fields like operations research, logistics, production planning, and finance, playing a crucial role in devising optimal solutions for complex decision-making scenarios. MILP solvers are developed to efficiently solve these problems and are integral to decision-making processes in various industrial and commercial applications.

The MILP solver in IBM ILOG CPLEX Optimization Suite~\cite{cplex} is a widely recognized solver since its introduction in 1991. Despite its long-standing reputation, some in the solver development community have cited CPLEX's seemingly weaker performance on the MIPLIB 2017\footnote{MIPLIB 2017 is a well-established set of 240 diverse optimization problems, representing a wide array of real-world scenarios.} benchmark~\cite{miplib}.  It is important to note that MIPLIB 2017, being a static and open benchmark set, allows solvers to be specifically tuned for these problems, a common practice among commercial solver teams in the recent decade. Interestingly, CPLEX has shown a relatively impartial stance towards such benchmark-specific tuning.

Our team works with CPLEX 12.9 released in March 2019\footnote{As of the time of this report, the latest CPLEX version is 22.1.1. However, version 12.9 remains more widely used in the community.}. Our research revealed that when CPLEX's configurations were adapted to our internal problem set, including MIPLIB 2017 as a subset, its performance was remarkably enhanced, surpassing other solvers in both solution time and the number of problems solved within a standard timeframe.

This report delves into the performance of the MindOpt Adapter for CPLEX (referred to as CPLEX Adapted) in comparison to the standard CPLEX configurations (referred to as CPLEX Default). CPLEX Adapted involves applying specific configurations based on the characteristics of the input problem. In our testing, CPLEX Adapted successfully solved 232 out of the 240 MIPLIB 2017 problems, surpassing the second-best solver. It also demonstrated superior performance in terms of the geometric mean of running times. Furthermore, in the set of 45 slightly pathological problems, CPLEX Adapted is the only solver that solved all the problems and still maintains a lead in terms of running times. In the 32-problem set designed for infeasibility detection, CPLEX Adapted correctly identified 31, exceeding the performance of other leading solvers and doing so more rapidly. Additionally, within a two-hour limit, CPLEX Adapted found the best incumbent solutions for 11 problems with previously less optimal solutions.

\section{MILP Benchmarks}
In the community, there are three datasets commonly used to evaluate the performance of MILP solvers~\cite{mittelmann_benchmarks}. In our experiments, we also used them to evaluate CPLEX Adapted, and achieved surprising results: CPLEX Adapted ranked first and solved more problems in all three datasets.

\begin{center}
\setlength{\tabcolsep}{4pt} 
\begin{tabular}{lrrrrrrrrrrrr}
\hline
          & CBC  & \textbf{GUROB} & COPT & SCIP & SCIPC & HiGHS & Matlb & SMOO & XSM & MDO & OPTV & \textbf{MDO4CPX} \\
\hline
unscal    & 1328 & \textbf{72.1}  & 126  & 888  & 727   & 720   & 2715  & 612  & 510   & 301   & 207  & \textbf{59.6} \\
scaled    & 18.4 & \textbf{1}     & 1.74 & 12.3 & 10.1  & 9.98  & 37.6  & 8.49 & 7.07  & 4.18  & 2.88 & \textbf{0.82} \\
solved    & 107  & \textbf{229}   & 212  & 137  & 152   & 159   & 73    & 163  & 172   & 196   & 202  & \textbf{232}  \\
\hline
\end{tabular}\\
{The MIPLIB2017 Benchmark}
\end{center}

\begin{center}
\setlength{\tabcolsep}{4pt} 
\begin{tabular}{lrrrrrrrrrr|r}
\hline
          & CBC  & COPT & GLPK & \textbf{GUROB} & HiGHS & MATLAB & MDOPT & SCIP & SCIPC & OPTV & \textbf{MDO4CPX} \\
\hline
unscal    & 7219  & 467  & 7317 &  \textbf{160}     & 3114  &  9760  &  2699 & 4733 & 3489  & 623 & \textbf{76.8}     \\
scaled    & 45.0 & 2.91 & 45.6 & \textbf{1}      & 19.4  & 60.9   & 16.8  & 28.8 & 22.5  & 3.89 &      \\
solved    & 5    & 41   & 6    & \textbf{43}     & 24    & 2      & 20    & 19   & 23    & 36   & \textbf{45}       \\
\hline
CPU & \multicolumn{10}{|c|}{Quad Intel Xeon E5-4657L, but using only 12 threads} & i7-11700K
\\
\hline
\end{tabular}\\
{The slightly pathological MILP Test (on different CPUs\footnote{
Unlike the other two datasets, Professor Hans Mittelmann evaluated solvers on the ``path'' dataset running in 12 threads on a platform with two or four Intel Xeon E5-4657L CPUs. %The Intel Xeon E5-4657L itself is a 12-core CPU, and to support a 4x expansion, we would need a four-socket Intel® Server Board S4600 series motherboard, which has been discontinued. Therefore, 
Due to the discontinuation of the motherboards compatible with this quad-CPU configuration, we were unable to replicate his testing environment. Consequently, we utilized the same platform of a single Intel i7-11700K CPU that was used for the other two datasets. Note that the results of the solvers may vary across the two platforms. Our configurations, obtained for i7-11700K, may not yield identical results on quad Xeon E5-4657L. Therefore, the performance results presented in this table from both CPUs should be considered for reference only and are not directly comparable.})}
\end{center}

\begin{center}
\setlength{\tabcolsep}{4pt} 
\begin{tabular}{lrrrrrrrrrrr}
\hline
          & CBC & COPT & \textbf{GUROB} & MATLAB & SCIP & SCIPC & HiGHS & MDOPT & OPTVER & \textbf{MDO4CPX} \\
\hline
scaled    & 22.5 & 1.47 & \textbf{1} & 44.5 & 9.22 & 8.18 & 7.81 & 7.19 & 3.00 & \textbf{0.71} \\
solved    & 20 & 30 & \textbf{30} & 16 & 25 & 26 & 26 & 27 & 28 & \textbf{31} \\
\hline
\end{tabular}\\
{The MILP Infeasibility Detection Test}
\end{center}

In the tables above, the results for solvers other than MDO4CPX\footnote{MDO4CPX is an abbreviation for MindOpt Adapter for CPLEX} are sourced from Professor Hans Mittelmann's benchmark website~\cite{mittelmann_benchmarks}, as of December 20, 2023. However, MDO4CPX was tested in our experimental setup, which closely matches his: a workstation equipped with an Intel i7-11700K CPU and 64GB of RAM.
Each test instance was given a maximum runtime of 7200, 3600, or 10800 seconds according to the testing requirements of different datasets. The MIP gap tolerance is set to 0, which means the solver needs to find and prove the optimal solution. The results are presented in tables where ``unscal'' and ``solved'' represent the unscaled shifted geometric means of the run times and the number of problems solved within the time limit, respectively.
He defines the shifted geometric mean (unscal) for $n$  nonnegative numbers $v_1, \ldots, v_n$ by the formula:
\[
\exp\left(\frac{\sum_{i=1}^{n} \ln(\max(1, v_i + s))}{n}\right) - s,
\]
where $v_i$ represents the Wallclock seconds of the $i$th test instance and the shift $s$ is set to 10. He computeds `scaled'' by divide the unscaled numbers by the reference (Gurob in these benchmarks) across the benchmarks. 

The three tables above highlights the relative performance of MindOpt Adapter for CPLEX against other solvers.

\subsection{Better Solutions Found for Unsolved Problems}

During these tests, on 11 specific problems from MIPLIB 2017, CPLEX Adapted identified better incumbent solutions within the allocated 2-hour time frame. We say these solutions are ``better'' in the sense that they are feasible and have objective values that are strictly lower than those of the previously recognized best-known solutions. Note that incumbent solutions are not necessarily optimal.

\begin{center}
\begin{tabular}{lcc}
\hline
Instance & New Incumbent Objective & Previous Best Objective \\ \hline
ns1690781 & -928.077871 & -927.0259 \\
neos-5221106-oparau & 54.65 & 55.54 \\
fhnw-binschedule0 & 16092 & 16122 \\
n3709 & 1206493 & 1207965 \\
neos-1420546 & 23005.21841 & 23011.81 \\
fhnw-schedule-pairb400 & -35.458406 & -35.45718 \\
n370b & 1220708 & 1225077 \\
lr1dr04vc05v17a-t360 & 252463.373469 & 252463.3952194264 \\
nsr8k & 18011349 & 18011358 \\
neos-1420790 & 3119.806 & 3120.439 \\
sct5 & -228.1412 & -228.1172304 \\
\hline
\end{tabular}
\end{center}

%\FloatBarrier  % Prevents floats from moving past this point

\subsection{Comparing Adapted to Default}
\label{sec:compare}
In the same environment as the previous experiments, we conducted a comparative analysis of CPLEX Default and CPLEX Adapted. This comparison was also based on the three datasets: "MIPLIB 2017 Instances," "MILP cases that are slightly pathological," and "Infeasibility Detection for MILP Problems."

\begin{center}
\begin{tabular}{lcccccc}
\hline
                 & \multicolumn{2}{c}{MIPLIB 2017} & \multicolumn{2}{c}{Pathological} & \multicolumn{2}{c}{Infeasible} \\
                 & \multicolumn{2}{c}{240 problems} & \multicolumn{2}{c}{45 problems} & \multicolumn{2}{c}{32 problems} \\ 
                 & unscal & solved & unscal & solved & unscal & detected \\
\hline
CPLEX Default  & 163    & 214    & 233    & 40     & 20     & 26     \\
CPLEX Adapted  & 59.6   & 232    & 76.8   & 45     & 5.6    & 31     \\
\hline
\end{tabular}
\end{center}

\begin{figure}[htbp] 
    \centering 
    \includegraphics[width=0.5\textwidth]{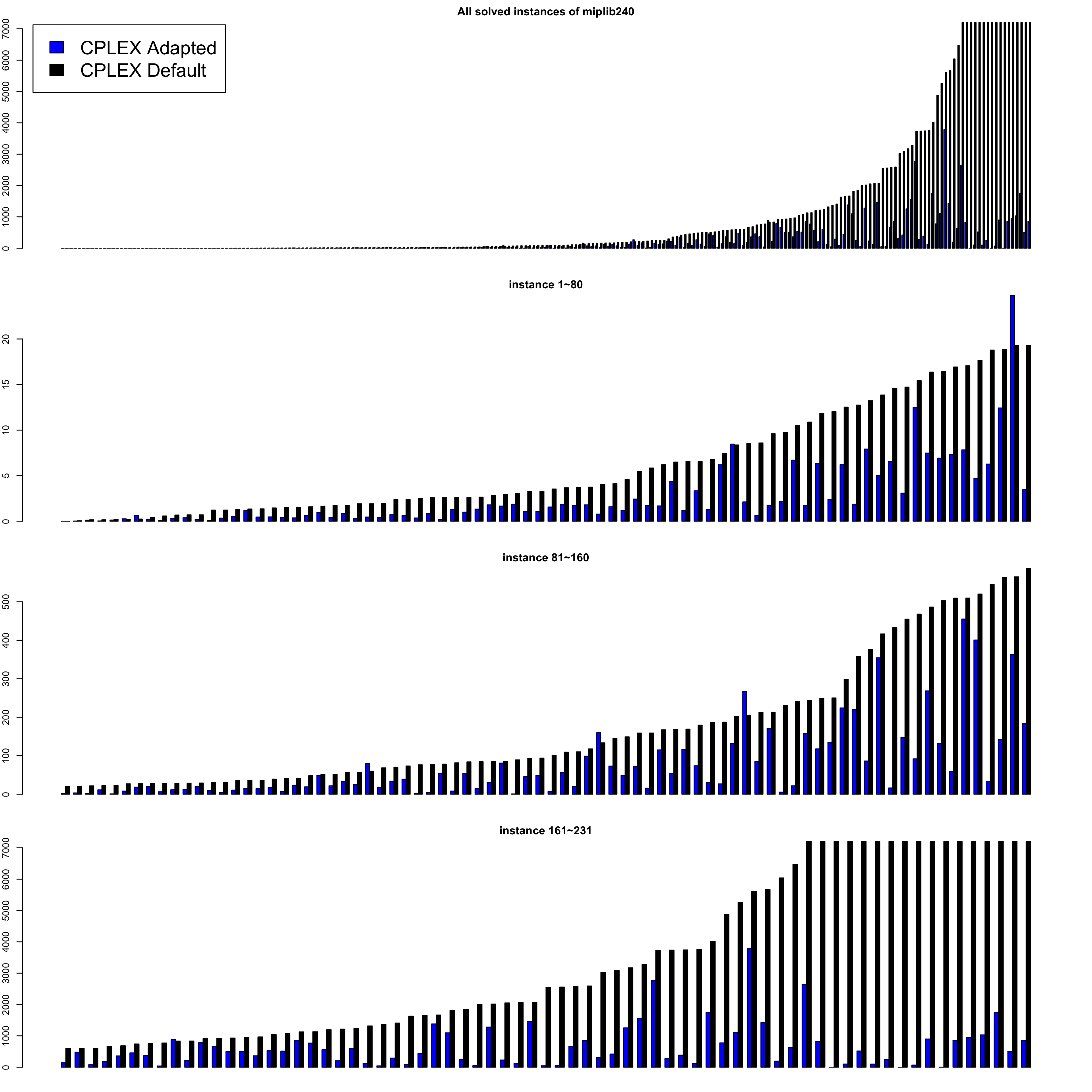} 
    \caption{Solution time distribution on the miplib240 dataset} 
    \label{fig:dist} 
\end{figure}

To provide a visual understanding of the differences between the two, Figure \ref{fig:dist} presents plots of the 240 times arranged in the ascending order according to CPLEX Default times. Most, but not all, problems see significant improvements with CPLEX Adapted. Notably, on problems where CPLEX Default faces difficulties, CPLEX Adapted tends to achieve more substantial speedups. Interestingly, on the problems that CPLEX Default failed to solve within the 2-hour time frame, CPLEX Adapted not only managed to solve them but also take relatively short times. This observation underscores the value of tuning configurations for problems, especially when they are challenging for default configurations.

\clearpage % 确保主体内容的所有表格和图片都已打印完毕

\section{CPLEX Configurations for the three MILP Benchmarks}
Table \ref{tab:configs} provides configurations that one can use to reproduce the evaluation results of CPLEX 12.9 with the three Benchmarks. Table \ref{tab:param} lists the parameter names as they appear in Configurations in order. 

Additionally, the problem neos-3046615-murg requires an additional set of warm-starting based on constraint aggregation; however, for the sake of brevity and simplicity, we chose not to elaborate this detail in the paper.

\begin{table}[!htbp]
  \centering
  \begin{tabular}{cc}
    \toprule
    Index & Parameter name \\
    \midrule  
1 & CPXPARAM\_MIP\_Cuts\_RLT \\
2 & CPXPARAM\_MIP\_Cuts\_MCFCut \\
3 & CPXPARAM\_Emphasis\_Numerical \\
4 & CPXPARAM\_MIP\_Strategy\_Dive \\
5 & CPXPARAM\_Preprocessing\_Dependency \\
6 & CPXPARAM\_MIP\_Limits\_GomoryCand \\
7 & CPXPARAM\_MIP\_Cuts\_Disjunctive \\
8 & CPXPARAM\_Preprocessing\_Folding \\
9 & CPXPARAM\_MIP\_Strategy\_SubAlgorithm \\
10 & CPXPARAM\_Preprocessing\_Relax \\
11 & CPXPARAM\_Simplex\_Crash \\
12 & CPXPARAM\_MIP\_Strategy\_Probe \\
13 & CPXPARAM\_MIP\_Cuts\_FlowCovers \\
14 & CPXPARAM\_MIP\_Cuts\_Covers \\
15 & CPXPARAM\_MIP\_Cuts\_Gomory \\
16 & CPXPARAM\_MIP\_Cuts\_Implied \\
17 & CPXPARAM\_Preprocessing\_Symmetry \\
18 & CPXPARAM\_MIP\_Cuts\_MIRCut \\
19 & CPXPARAM\_MIP\_Strategy\_VariableSelect \\
20 & CPXPARAM\_MIP\_Cuts\_LocalImplied \\
21 & CPXPARAM\_MIP\_Cuts\_ZeroHalfCut \\
22 & CPXPARAM\_Preprocessing\_Dual \\
23 & CPXPARAM\_MIP\_Cuts\_BQP \\
24 & CPXPARAM\_Preprocessing\_CoeffReduce \\
25 & CPXPARAM\_MIP\_Strategy\_FPHeur \\
26 & CPXPARAM\_MIP\_Limits\_AggForCut \\
27 & CPXPARAM\_MIP\_Strategy\_StartAlgorithm \\
28 & CPXPARAM\_MIP\_Strategy\_Search \\
29 & CPXPARAM\_MIP\_Cuts\_Cliques \\
30 & CPXPARAM\_MIP\_SubMIP\_StartAlg \\
31 & CPXPARAM\_Preprocessing\_Reduce \\
32 & CPXPARAM\_MIP\_Limits\_CutsFactor \\
33 & CPXPARAM\_Preprocessing\_RepeatPresolve \\
34 & CPXPARAM\_Threads \\
35 & CPXPARAM\_MIP\_SubMIP\_SubAlg \\
36 & CPXPARAM\_Preprocessing\_BoundStrength \\
37 & CPXPARAM\_MIP\_Strategy\_NodeSelect \\
38 & CPXPARAM\_MIP\_Strategy\_PresolveNode \\
39 & CPXPARAM\_MIP\_Strategy\_Branch \\
40 & CPXPARAM\_MIP\_Cuts\_PathCut \\
41 & CPXPARAM\_MIP\_Cuts\_LiftProj \\
42 & CPXPARAM\_Emphasis\_MIP \\
43 & CPXPARAM\_Preprocessing\_Linear \\
44 & CPXPARAM\_MIP\_Strategy\_RINSHeur \\
45 & CPXPARAM\_MIP\_Cuts\_GUBCovers \\
46 & CPXPARAM\_MIP\_Tolerances\_MIPGap \\
47 & CPXPARAM\_Advance \\
    \bottomrule  
  \end{tabular}
  \caption{Used configuration parameters}
  \label{tab:param}
\end{table}

\begin{longtable}[htbp]{cccc}
\caption{Configurations and solution times with the MILP Benchmarks} \label{tab:configs} \\
\hline
Instance & Adapted & Default & Configurations \\ \hline
\endfirsthead

\multicolumn{4}{c}
{{\bfseries \tablename\ \thetable{} -- continued from previous page}} \\
\hline
Instance & Adapted & Default & Configuration \\ \hline
\endhead

\hline %%\multicolumn{4}{r}{{Continued on next page}} \\
\endfoot

\hline
\endlastfoot

% Your table data starts here
nursesched-sprint02 & 1.67 & 2.97 & ,,,,0,,-1,,,0,,,-1,1,,2,0,-1,-1,,1,-1,,2,,,,,-1,,,,,8,,0,0,1,1,,,,0,2,-1,,2 \\ 
sct2 & 4.36 & 6.49 & ,,,,3,,3,,,,,3,2,-1,-1,2,,,-1,,2,1,,0,2,,6,,-1,,,,,8,,,0,2,,2,,,,-1,-1,,0 \\ 
neos-950242 & 2.45 & 19.71 & ,,,1,3,,,,4,,,-1,2,,,2,5,1,4,,,1,,0,-1,,5,,2,,,,0,4,,1,2,1,-1,,,2,,-1,-1,,2 \\ 
ns1830653 & 7.92 & 13.23 & ,,,,,,,,,,,,,,,,,,,,,,,2,,,,,,,,,0,8,,,,,,,,1,,,,, \\ 
dano3\_3 & 6.2 & 12.53 & ,,,2,,,-1,,,0,,,2,2,1,-1,,2,-1,,2,,,1,,,6,,,,1,,,8,,0,,2,,-1,,,0,-1,-1,, \\ 
sp150x300d & 0.03 & 0.16 & ,,,1,0,,3,,,1,,3,2,2,,2,5,-1,,,-1,-1,,2,,,3,,2,,,,,8,,,3,2,,-1,,,,5,,,0 \\ 
savsched1 & 40.48 & 772.5 & ,,,3,,,,,,,,-1,2,3,2,-1,,2,-1,,2,1,,2,-1,,1,,3,,,,3,8,,,,1,-1,2,,1,,4,-1,, \\ 
pg5\_34 & 1.56 & 3.54 & ,,,,0,,-1,,,,,,-1,,-1,1,,1,,,1,,,2,,,,,-1,,,,,8,,,,1,-1,,,,,-1,-1,, \\ 
control30-5-10-4 & 0.13 & 4.53 & ,,,,,,,,,,,,,,,,,,,,,,,,,,,,,,,,,,,,,,,,,,,,,, \\ 
roi2alpha3n4 & 11.29 & 22.39 & ,,,1,,,-1,,,,,,1,2,-1,-1,,2,4,,1,-1,,2,-1,,6,,-1,,,,3,8,,0,0,2,1,2,,1,,,,, \\ 
neos-3627168-kasai & 54.52 & 168.12 & ,,,,0,,,,,,,,,1,,,,,,,-1,1,,2,,,,,,,,,,8,,,,1,1,,,,,-1,,,2 \\ 
drayage-100-23 & 0.13 & 0.21 & ,,,,,,3,,,,,-1,-1,2,-1,2,5,-1,-1,,2,1,,2,-1,,6,,3,,,,,8,,1,3,2,1,-1,,1,,-1,2,,0 \\ 
neos17 & 0.44 & 1.5 & ,,,,,,,,,,,,,,,,,,,,,,,2,,,,,,,,,,8,,,,,,,,,,,,,0 \\ 
cryptanalysiskb128n5obj16 & 48.63 & 2003.99 & ,,,,3,,1,,,1,,3,-1,3,2,1,5,1,4,,-1,1,,0,,,3,,-1,,0,,,8,,0,,2,1,,,2,,5,2,,0 \\ 
ns2319914 & 46.02 & 1809.59 & ,,,,,,-1,,5,,,-1,-1,3,-1,2,,-1,-1,,-1,-1,,2,-1,,6,,3,,0,,,11,,1,0,-1,-1,2,,,0,-1,2,,0 \\ 
icir97\_tension & 56.72 & 109.38 & ,,,,,,,,,,,,,,,,,,,,,,,2,,,,,,,,,,8,,,,,,-1,,,,,,, \\ 
ns2017839 & 5.4 & 9.28 & ,,,1,,,,,,0,,,-1,-1,2,1,,2,1,,,-1,,2,-1,,,,-1,,0,,,10,,,3,,1,-1,,4,,-1,,,0 \\ 
neos-787933 & 0.63 & 0.24 & ,,,2,0,,-1,,,,,3,-1,2,,,1,1,2,,,,,1,,,,,-1,,,,2,8,,1,2,-1,1,,,2,,5,2,, \\ 
ns43503 & 5.98 & 6.11 & ,,,,,,,,,,,,,,,,,,,,,,,2,,,,,,,,,,8,,,,,,,,,,,,, \\ 
neos-5052403-cygnet & 122.54 & 1316.05 & ,,,3,2,,-1,,,0,,3,-1,3,2,1,5,2,,,-1,,,2,,,2,,3,,0,,1,8,,,0,-1,,1,,,0,-1,1,, \\ 
mas74 & 14.45 & 84.45 & ,,,2,0,,,,2,1,,3,2,1,,1,5,-1,2,,1,1,,0,1,,6,,-1,,2,,0,8,,,3,,-1,1,,,,-1,-1,,0 \\ 
piperout-08 & 0.63 & 1.59 & ,,,,0,,2,,,,,,-1,1,1,,0,-1,-1,,1,-1,,2,,,1,,,,,,1,8,,,,-1,-1,-1,,2,0,-1,2,, \\ 
mzzv11 & 1.19 & 6.55 & ,,,1,2,,-1,,,,,2,,-1,,-1,,1,,,-1,,,2,-1,,1,,2,,2,,3,8,,0,,,1,,,1,,4,1,,0 \\ 
satellites2-40 & 81.01 & 607.62 & ,,,,,,1,,,1,,1,,1,,,4,2,2,,,-1,,0,2,,,,2,,,,2,8,,,3,2,1,2,,4,,1,2,,0 \\ 
neos-1354092 & 184.29 & 586.4 & ,,,3,,,3,,,,,2,-1,-1,2,2,,-1,4,,1,-1,,2,-1,,,,-1,,,,,8,,,3,2,-1,1,,4,0,-1,,,0 \\ 
flugplinf & 0 & 0.02 & ,,,1,,,,,,1,,-1,-1,3,2,-1,,2,4,,,,,0,-1,,1,,3,,,,3,1,,1,0,2,-1,2,,4,0,-1,,,2 \\ 
sp98ar & 98.95 & 117.8 & ,,,,,,,,,,,,-1,,-1,,,,,,,1,,2,,,,,,,,,,8,,0,,,-1,-1,,,,,,, \\ 
rmatr200-p5 & 1282.41 & 2015.28 & ,,,2,1,,-1,,,1,,-1,2,3,,1,4,2,-1,,2,-1,,0,2,,2,,3,,0,,,8,,0,3,-1,-1,2,,1,0,-1,2,,2 \\ 
peg-solitaire-a3 & 21.89 & 241.37 & ,,,,1,,2,,4,1,,3,1,2,,2,4,-1,3,,1,,,1,-1,,1,,1,,,,,8,,,,1,,-1,,,0,,-1,,0 \\ 
neos-631710 & 54.93 & 78.09 & ,,,,2,,2,,1,0,,,-1,,2,1,4,1,2,,-1,,,0,2,,3,,,,,,,8,,0,0,1,1,,,4,,2,2,, \\ 
supportcase26 & 16.26 & 432.82 & ,,,3,3,,2,,2,0,,,2,-1,2,-1,5,-1,3,,-1,,,0,2,,,,1,,0,,,8,,1,,-1,1,,,4,,,-1,,0 \\ 
reblock115 & 769.82 & 1131.37 & ,,,,,,,,,,,,,,,,,,,,,,,1,,,,,,,,,,8,,,,-1,1,-1,,,,-1,-1,,0 \\ 
fhnw-binpack4-48 & 2.65 & 7200 & ,,,1,,,,,,0,,-1,2,3,2,2,5,2,4,,2,-1,,0,1,,6,,2,,0,,0,8,,,2,1,1,,,2,0,-1,1,,2 \\ 
neos-4413714-turia & 7.05 & 40.43 & ,,,,2,,1,,,,,-1,,3,-1,-1,0,-1,-1,,2,,,0,-1,,2,,-1,,,,,8,,0,0,1,-1,2,,,0,-1,2,,2 \\ 
enlight9 & 0 & 0.01 & ,,,1,1,,-1,,4,1,,3,2,3,-1,-1,0,1,,,2,-1,,1,-1,,1,,2,,1,,2,5,,0,0,2,1,2,,3,,3,2,,2 \\ 
uccase9 & 782 & 909.94 & ,,,,,,,,,,,,,,,,0,,,,,,,2,,,,,,,,,,8,,,,,,,,,,,,, \\ 
rocI-4-11 & 6.35 & 11.84 & ,,,,,,,,,0,,,,1,,,1,,,,,,,1,-1,,1,,-1,,,,,8,,1,,,1,,,,,-1,,, \\ 
neos-2656603-coxs & 0.28 & 1.73 & ,,,,,,1,,2,,,-1,-1,-1,,,,2,,,-1,,,2,,,1,,,,,,,8,,,2,,,1,,1,,-1,-1,, \\ 
neos-5107597-kakapo & 11.67 & 28.24 & ,,,1,,,1,,,,,1,-1,-1,1,-1,,1,,,,-1,,2,-1,,,,1,,1,,1,8,,,,1,,1,,1,0,,2,,0 \\ 
comp21-2idx & 2648 & 7200 & ,,,,,,-1,,1,1,,-1,,3,-1,,4,-1,4,,2,,,2,-1,,6,,1,,,,,8,,1,3,,1,2,,,0,-1,-1,,2 \\ 
traininstance6 & 188 & 5.49 & ,,,,,,1,,,,,,1,,,-1,,1,,,,,,2,,,,,,,,,,8,,0,,,,-1,,,,,,, \\ 
neos-3656078-kumeu & 900.75 & 7200 & ,,,2,0,,,,,0,,,1,,,,,,,,1,,,2,,,,,,,,,,8,,,,1,-1,,,,,,,,2 \\ 
gmu-35-40 & 85.73 & 212.53 & ,,,,,,,,,,,1,-1,1,1,-1,,1,,,1,1,,1,1,,,,1,,,,,8,,,0,-1,1,1,,,,,-1,,0 \\ 
h80x6320d & 1.33 & 2.64 & ,,,2,,,1,,,1,,2,2,2,-1,-1,2,-1,2,,2,-1,,0,-1,,1,,3,,,,3,8,,,0,1,-1,-1,,,0,5,2,,0 \\ 
irish-electricity & 862 & 1126.24 & ,,,,,,1,,,,,1,,-1,,,,,,,1,,,2,,,,,-1,,,,,8,,,,1,,-1,,,0,,-1,,2 \\ 
neos-3754480-nidda & 386.18 & 3742.63 & ,,,,1,,-1,,,,,-1,-1,-1,-1,2,2,-1,4,,-1,1,,0,2,,6,,-1,,0,,0,8,,,0,-1,1,,,4,,-1,1,, \\ 
bnatt500 & 1116.9 & 5258.16 & ,,,1,,,,,2,,,,-1,,,1,,,,,,,,1,,,,,1,,,,0,8,,1,,,,,,,,,,,0 \\ 
cbs-cta & 0.19 & 0.7 & ,,,3,3,,3,,5,0,,-1,2,3,,,0,-1,3,,-1,,,0,1,,,,,,,,0,8,,0,2,-1,-1,,,3,0,,,,2 \\ 
lectsched-5-obj & 194.89 & 6039.37 & ,,,,,,,,1,0,,,,,,,0,,,,,,,2,,,1,,,,,,,8,,,,,,,,,0,,,,2 \\ 
nu25-pr12 & 0.41 & 1.96 & ,,,,0,,1,,2,0,,2,-1,,1,-1,1,2,2,,-1,,,2,,,,,1,,0,,2,8,,,3,2,1,2,,4,,,-1,, \\ 
neos-957323 & 2.14 & 9.75 & ,,,,0,,2,,,,,-1,2,3,-1,1,4,2,3,,2,-1,,0,-1,,,,-1,,,,1,8,,0,2,,-1,-1,,4,,,-1,,2 \\ 
csched007 & 73.1 & 145.24 & ,,,,,,,,,1,,,1,-1,,-1,3,,,,-1,1,,2,1,,1,,1,,,,3,8,,0,2,1,,,,,,-1,-1,,0 \\ 
mzzv42z & 0.79 & 4.04 & ,,,,3,,,,2,,,2,-1,-1,1,-1,1,1,3,,2,-1,,2,,,1,,2,,0,,0,8,,,0,2,-1,,,1,,1,-1,, \\ 
comp07-2idx & 8.17 & 27.16 & ,,,1,2,,-1,,1,0,,1,-1,3,-1,-1,5,-1,4,,2,1,,0,1,,1,,,,,,2,8,,,,,-1,2,,,0,3,-1,,0 \\ 
misc05inf & 0.01 & 0.04 & ,,,1,,,-1,,,0,,-1,,3,1,1,5,,,,2,-1,,1,,,2,,2,,,,0,1,,,3,,-1,-1,,2,,-1,2,, \\ 
rocII-5-11 & 86.3 & 375.66 & ,,,1,0,,,,,0,,,1,,,-1,1,-1,,,,,,1,-1,,,,1,,,,,8,,0,,,,-1,,,,,1,,2 \\ 
supportcase42 & 115.17 & 167.39 & ,,,3,3,,2,,,,,2,2,1,1,-1,5,-1,3,,-1,-1,,0,1,,1,,-1,,,,2,8,,0,,1,,1,,3,0,4,2,,2 \\ 
ns4976508 & 2030.25 & 10800.21 & ,,,3,1,,2,,2,0,,1,,,2,,3,-1,2,,,,,2,-1,,3,,1,,2,,1,6,,,3,,,,,2,0,3,-1,, \\ 
timtab1 & 12.84 & 28.67 & ,,,,,,,,,,,-1,,-1,,-1,,1,,,,,,2,-1,,,,,,,,,8,,,,1,,,,,,,1,, \\ 
gen-ip054 & 33.88 & 70.45 & ,,,1,2,,2,,2,,,-1,,3,2,1,3,1,2,,2,1,,2,2,,2,,1,,2,,2,8,,0,0,,1,-1,,4,,,,,2 \\ 
no-ip-64999 & 1.14 & 10.12 & ,,,2,3,,-1,,,1,,,-1,2,2,,4,2,4,,,1,,0,2,,6,,3,,,,3,8,,,0,1,-1,1,,1,0,-1,1,,0 \\ 
neos-5104907-jarama & 947.94 & 7200 & ,,,2,0,,,,2,0,,2,1,1,,,1,-1,4,,,,,0,-1,,2,,1,,2,,1,7,,0,,,,2,,2,0,2,,, \\ 
piperout-27 & 0.61 & 2.37 & ,,,,3,,-1,,,1,,-1,2,3,-1,-1,4,2,4,,,-1,,2,-1,,,,-1,,,,0,8,,,2,-1,,1,,1,,,2,, \\ 
neos-5093327-huahum & 882.11 & 833.2 & ,,,,,,,,,,,,,,,,,,,,,,,2,,,,,,,,,,8,,,,,,,,,,,,, \\ 
neos-3988577-wolgan & 3.54 & 3603.05 & ,,,1,3,,-1,,,1,,3,-1,1,,1,4,,,,1,-1,,0,,,2,,3,,1,,1,8,,0,0,-1,-1,,,3,,3,-1,,0 \\ 
trento1 & 33.79 & 56.57 & ,,,,,,1,,,,,,-1,-1,1,,,,,,,,,2,-1,,,,,,,,0,8,,,,,,,,,,,,,0 \\ 
neos-2657525-crna & 254.15 & 7200 & ,,,,,,3,,,0,,-1,2,-1,-1,1,2,2,4,,2,1,,0,2,,6,,3,,0,,0,8,,1,2,2,-1,-1,,2,0,-1,2,,0 \\ 
neos-2075418-temuka & 99.4 & 3600.3 & ,,,,2,,,,,1,,3,-1,3,2,2,,-1,-1,,-1,-1,,0,-1,,6,,-1,,,,3,8,,1,2,2,1,1,,4,0,5,-1,, \\ 
dws008-01 & 1380.82 & 1665.96 & ,,,,,,,,,,,,,,,,,,,,,,,2,,,,,,,,,,8,,,,,,,,,,,,, \\ 
enlight11 & 0 & 0.01 & ,,,2,2,,3,,3,0,,-1,,-1,2,,2,,-1,,,-1,,0,-1,,,,-1,,,,0,1,,,3,,,1,,2,0,4,,, \\ 
cost266-UUE & 267.62 & 205.39 & ,,,,,,,,,,,,-1,,,,,,,,,,,2,,,,,,,,,,8,,,,,,,,,,,,, \\ 
bab6 & 365.48 & 756.58 & ,,,3,,,-1,,,,,3,-1,,2,-1,0,-1,-1,,,1,,2,2,,1,,2,,,,,4,,,0,-1,,2,,,0,,2,, \\ 
n2seq36q & 0.66 & 8.59 & ,,,3,,,2,,1,,,-1,-1,-1,2,2,5,2,-1,,-1,-1,,2,-1,,1,,,,2,,3,8,,1,3,-1,-1,-1,,4,0,-1,2,,2 \\ 
ns3337549 & 3636.57 & 10803.5 & ,,,3,3,,-1,,2,0,,,2,-1,1,-1,1,-1,,,2,,,2,2,,3,,-1,,1,,3,8,,1,3,-1,1,1,,2,0,,2,,2 \\ 
neos-3046615-murg & 0.85 & 7200 & ,,,,0,,-1,,,1,,-1,-1,1,2,1,0,-1,,,1,1,,1,2,,,,3,,,,0,8,,1,0,2,,,,,,,,,2 \\ 
glass-sc & 363.32 & 564.76 & ,,,,,,1,,,,,-1,1,-1,-1,1,,1,,,1,-1,,2,,,,,-1,,2,,0,8,,0,2,-1,1,-1,,,,,1,,0 \\ 
blp-ic98 & 18.37 & 27.48 & ,,,,,,,,,,,,,,,,,,,,,,,2,,,,,,,,,,8,,,,,,,,,0,,,, \\ 
fast0507 & 7.32 & 16.93 & ,,,,,,-1,,,0,,-1,-1,-1,-1,2,0,-1,,,-1,-1,,1,1,,,,,,,,,8,,,0,2,1,,,,,-1,,,0 \\ 
supportcase18 & 3.38 & 20.92 & ,,,2,,,1,,,1,,2,,1,,-1,1,,3,,1,,,1,,,2,,,,,,,8,,0,0,,1,-1,,,0,1,-1,, \\ 
ns1644855 & 90.56 & 1628.69 & ,,,2,0,,-1,,,0,,3,1,1,1,-1,4,-1,4,,2,,,2,2,,4,,1,,2,,3,7,,,0,1,-1,-1,,4,0,3,1,, \\ 
neos-911970 & 0.71 & 89.29 & ,,,,3,,-1,,,,,-1,2,,1,-1,5,2,4,,-1,-1,,0,2,,,,-1,,1,,,8,,1,2,-1,1,2,,,,-1,,,0 \\ 
neos-933966 & 1.86 & 3.68 & ,,,1,,,-1,,,1,,-1,2,2,-1,-1,5,2,-1,,2,-1,,2,2,,4,,-1,,,,1,8,,1,3,,1,2,,4,0,5,-1,,0 \\ 
decomp2 & 0.07 & 1.23 & ,,,2,3,,1,,,0,,,,2,1,2,5,1,3,,-1,-1,,1,,,2,,,,2,,,8,,0,3,1,-1,2,,1,,1,,,2 \\ 
momentum1 & 672.38 & 2578.36 & ,,,,,,,,1,,,,,,,,,,,,,,,2,,,1,,,,,,0,8,,,,,,,,,,,,, \\ 
glass4 & 10.73 & 35.39 & ,,,3,2,,-1,,,,,3,2,-1,2,-1,,2,4,,-1,-1,,0,2,,,,-1,,,,,8,,,0,2,1,1,,4,0,4,-1,,0 \\ 
gmu-35-50 & 558.37 & 1197.92 & ,,,,,,,,,,,,,,,,,,,,,,,2,,,,,,,,,,8,,,,,,,,,,,,, \\ 
ns3633010 & 23.69 & 53.91 & ,,,,,,,,,,,,,,,-1,,,,,,,,2,,,,,,,,,,8,,,,,,,,,,,,, \\ 
blp-ar98 & 54.48 & 84.13 & ,,,,0,,-1,,,,,,,-1,-1,-1,0,1,,,1,1,,2,1,,,,-1,,,,,8,,,,,,,,,,-1,-1,, \\ 
supportcase10 & 630.04 & 6474.74 & ,,,,,,-1,,4,,,-1,-1,1,1,2,,-1,4,,,1,,2,2,,,,3,,,,0,8,,1,2,2,1,-1,,1,0,,,,0 \\ 
ex9 & 1.16 & 1.34 & ,,,2,,,3,,5,0,,3,2,-1,2,2,,,3,,-1,-1,,0,2,,4,,3,,,,1,1,,0,3,2,1,-1,,,,5,2,,0 \\ 
supportcase40 & 48.78 & 149.17 & ,,,,3,,,,,,,-1,1,3,2,-1,3,,2,,1,-1,,2,-1,,2,,2,,,,2,8,,0,2,-1,1,2,,1,0,-1,1,, \\ 
seymour1 & 6.19 & 7.45 & ,,,,,,-1,,,,,-1,1,-1,,-1,5,1,2,,2,1,,2,-1,,,,-1,,0,,3,8,,,3,2,-1,-1,,4,0,-1,2,,2 \\ 
map10 & 49.1 & 51.34 & ,,,,,,-1,,,,,-1,1,,-1,-1,,-1,,,,-1,,2,-1,,,,1,,,,,8,,,0,-1,-1,-1,,,,-1,,,2 \\ 
snp-02-004-104 & 218.06 & 834.7 & ,,,,1,,,,,0,,1,-1,,,,,,,,1,1,,1,-1,,,,,,,,0,8,,1,,,-1,,,,0,-1,1,,0 \\ 
atlanta-ip & 455.17 & 509.59 & ,,,2,,,,,,1,,-1,-1,,,-1,0,-1,,,,,,1,-1,,1,,,,,,,8,,,3,,1,,,,,-1,2,,2 \\ 
p2m2p1m1p0n100 & 0.18 & 0.93 & ,,,1,,,-1,,,,,,1,1,,-1,0,-1,,,1,1,,2,1,,,,,,,,,8,,,2,2,-1,1,,,,-1,,,2 \\ 
neos-849702 & 5.71 & 24.12 & ,,,,3,,3,,,,,,-1,2,2,2,,-1,4,,2,,,2,1,,,,-1,,0,,,12,,,0,,1,1,,,,-1,2,, \\ 
ns1116954 & 1.59 & 4.12 & ,,,3,1,,1,,4,1,,3,2,1,,-1,1,,4,,,-1,,2,,,2,,2,,1,,0,8,,0,3,-1,-1,1,,4,,5,1,,2 \\ 
mik-250-20-75-4 & 0.36 & 2.53 & ,,,3,3,,3,,,1,,3,2,-1,,-1,,2,2,,-1,1,,0,2,,6,,3,,,,3,8,,1,2,-1,,,,2,,-1,1,,0 \\ 
ns1954122 & 33.04 & 139.07 & ,,,3,3,,3,,,1,,-1,-1,1,-1,2,,2,-1,,2,-1,,0,,,5,,-1,,2,,1,3,,1,0,2,,-1,,,0,-1,-1,,0 \\ 
ns3134812 & 44.94 & 1105.22 & ,,,,,,-1,,4,0,,1,,1,2,-1,,,,,2,1,,0,,,3,,,,,,0,11,,1,0,2,-1,1,,,,5,,,2 \\ 
p200x1188c & 0.07 & 0.58 & ,,,,,,1,,2,0,,-1,,-1,2,1,1,-1,1,,-1,1,,1,,,,,-1,,,,,8,,,0,2,-1,1,,3,,-1,,,0 \\ 
neos8 & 0.26 & 0.22 & ,,,3,2,,-1,,,1,,-1,-1,,-1,,2,2,,,1,1,,1,-1,,3,,-1,,1,,1,8,,,2,,1,,,,,5,,, \\ 
neos-1171448 & 1.08 & 3.26 & ,,,3,1,,1,,1,,,2,-1,1,-1,,3,-1,4,,-1,-1,,0,1,,2,,,,,,0,8,,1,3,-1,-1,-1,,1,0,3,1,,0 \\ 
ns5013590 & 294.39 & 484.17 & ,,,1,,,-1,,,0,,,-1,-1,1,-1,,-1,,,-1,,,2,-1,,,,-1,,,,,8,,,0,,1,1,,,,-1,-1,,2 \\ 
ns4165869 & 716.53 & 5921.01 & ,,,3,,,-1,,,1,,2,2,-1,2,2,,2,-1,,2,1,,2,2,,6,,-1,,,,3,12,,,0,2,-1,-1,,,,-1,-1,,2 \\ 
sorrell3 & 530.16 & 1038.4 & ,,,,0,,-1,,,0,,,1,-1,,1,,,,,,,,2,-1,,2,,2,,,,0,8,,,,,,1,,,0,-1,1,, \\ 
chrom\_512 & 7.83 & 78.72 & ,,,3,,,,,,,,-1,2,,-1,-1,5,1,3,,-1,-1,,2,-1,,1,,3,,,,0,8,,1,3,1,1,1,,1,,5,1,,2 \\ 
ns2350781 & 480.78 & 290.21 & ,,,,,,,,,,,,,,,,,,,,,,,2,,,,,,,,,,8,,,,,,,,,,,,, \\ 
nursesched-medium-hint03 & 2775 & 3730.12 & ,,,,,,-1,,2,,,1,2,-1,2,-1,5,2,,,,-1,,1,2,,,,1,,1,,3,8,,0,2,2,1,1,,3,,5,2,,0 \\ 
physiciansched6-2 & 1.18 & 4.57 & ,,,3,,,-1,,,,,3,2,3,2,-1,,-1,4,,-1,1,,2,1,,6,,-1,,,,0,8,,,0,1,,2,,,,5,-1,,2 \\ 
neos-4722843-widden & 4.05 & 31.3 & ,,,1,,,,,,,,-1,-1,-1,,1,4,2,3,,-1,1,,0,1,,5,,3,,,,,8,,0,2,1,-1,,,1,,5,2,,0 \\ 
chromaticindex512-7 & 31.02 & 85.57 & ,,,,,,-1,,2,,,,-1,-1,,2,,2,,,1,,,2,-1,,1,,-1,,,,,8,,,0,-1,,,,,0,-1,,,2 \\ 
cod105 & 6.57 & 14.59 & ,,,2,0,,-1,,,1,,,1,3,1,2,,1,,,2,-1,,0,-1,,,,-1,,,,0,8,,0,0,1,,2,,1,0,-1,1,,0 \\ 
ns3974959 & 49.36 & 10811.37 & ,,,,,,,,,,,,,,,,,,,,,,,2,,,,,,,,,,8,,,,,,,,,,,,, \\ 
chrom\_256 & 2.56 & 11.75 & ,,,1,0,,2,,,1,,-1,-1,2,-1,2,,2,2,,-1,-1,,2,-1,,1,,-1,,,,0,10,,,0,1,-1,,,1,0,,,,2 \\ 
neos-4954672-berkel & 1255.83 & 3172.8 & ,,,,,,,,,,,-1,,,,,,,,,,,,2,-1,,,,,,,,,8,,,,,,1,,,,,,, \\ 
rail02 & 1737.1 & 7200 & ,,,,3,,2,,,,,-1,-1,3,-1,-1,5,-1,3,,-1,1,,2,1,,6,,3,,,,2,8,,0,,2,-1,2,,4,0,2,-1,,0 \\ 
neos-848589 & 147.79 & 454.8 & ,,,2,,,1,,,,,,2,-1,1,1,4,,-1,,,,,0,,,,,1,,,,3,8,,,,2,-1,2,,,,-1,2,,2 \\ 
neos-3135526-osun & 0.01 & 0.05 & ,,,,2,,,,1,1,,2,-1,2,2,-1,2,-1,4,,-1,1,,1,2,,4,,3,,2,,0,5,,0,2,2,-1,2,,,,-1,,,2 \\ 
neos-3218348-suir & 0.28 & 0.45 & ,,,1,0,,,,,0,,1,,-1,1,-1,0,1,,,1,,,1,-1,,,,,,,,0,8,,,2,2,,1,,1,,-1,,,0 \\ 
ns29903 & 660.46 & 875.25 & ,,,,,,2,,,,,1,1,2,1,1,3,,3,,-1,-1,,2,-1,,,,-1,,,,3,12,,,,-1,1,-1,,,,,-1,,0 \\ 
ns2034125 & 86.62 & 744.59 & ,,,3,,,3,,,1,,3,2,1,2,2,,1,-1,,,1,,0,,,6,,-1,,,,,12,,,3,2,,2,,,,-1,-1,,0 \\ 
mad & 231.4 & 2050.89 & ,,,,0,,-1,,,1,,-1,-1,1,2,1,0,-1,,,1,1,,1,2,,,,3,,,,0,8,,1,0,2,,,,,,,,,2 \\ 
ponderthis0517-inf & 0.02 & 3600.05 & ,,,3,,,-1,,,1,,-1,-1,,-1,2,3,2,4,,2,-1,,2,-1,,5,,3,,0,,,1,,1,0,1,,2,,2,,-1,2,, \\ 
supportcase6 & 8.32 & 81.46 & ,,,3,1,,-1,,2,1,,-1,1,1,-1,1,,-1,3,,2,-1,,1,-1,,5,,-1,,0,,3,8,,1,3,1,,2,,1,0,5,-1,,2 \\ 
pg & 0.46 & 1.91 & ,,,3,3,,-1,,,,,1,-1,,-1,2,5,-1,1,,2,,,0,2,,,,-1,,0,,,8,,0,0,,1,2,,,,-1,2,,2 \\ 
neos-1171737 & 1.8 & 2.85 & ,,,2,0,,,,,1,,3,-1,-1,-1,-1,0,-1,3,,-1,1,,0,,,3,,2,,,,3,8,,1,3,1,1,2,,1,,4,1,,2 \\ 
neos-5188808-nattai & 145.41 & 596.12 & ,,,,,,,,,0,,-1,-1,-1,,,,-1,,,-1,1,,2,-1,,,,-1,,,,,8,,,2,,-1,-1,,,,-1,1,, \\ 
ns2844866 & 0.39 & 1.14 & ,,,3,3,,3,,,,,3,2,3,2,2,5,-1,4,,1,,,2,2,,,,-1,,0,,,12,,,0,1,1,-1,,,,-1,-1,,0 \\ 
no-ip-65059 & 0.22 & 11.82 & ,,,2,2,,,,,0,,,-1,2,-1,-1,4,,2,,,1,,2,-1,,1,,1,,,,3,8,,,3,2,1,1,,4,,,-1,, \\ 
neos-960392 & 1.87 & 12.75 & ,,,,1,,,,1,0,,,-1,3,-1,2,5,-1,4,,,-1,,2,1,,3,,1,,1,,3,8,,0,,2,-1,,,,0,2,-1,,0 \\ 
ic97\_potential & 293.12 & 1410.53 & ,,,,3,,3,,,0,,-1,1,-1,1,1,,-1,,,1,,,2,,,6,,3,,,,1,8,,1,2,1,1,2,,,0,-1,-1,, \\ 
tbfp-network & 19.87 & 110.11 & ,,,,1,,1,,,,,-1,-1,-1,2,1,0,1,-1,,1,,,2,2,,,,-1,,,,0,8,,,3,,1,,,,0,-1,-1,,0 \\ 
rococoB10-011000 & 400.74 & 520.04 & ,,,,,,,,,,,,,,,,,,,,,,,2,,,,,,,,,,8,,,,,,,,,,,,, \\ 
wachplan & 116.46 & 169.14 & ,,,2,0,,2,,2,1,,3,,2,,-1,4,,1,,,-1,,0,1,,,,2,,,,0,6,,1,2,2,1,-1,,2,0,2,-1,,0 \\ 
square47 & 495.95 & 931.12 & ,,,1,0,,,,,0,,-1,2,,-1,2,,,,,,-1,,1,-1,,,,1,,,,0,8,,,,1,-1,1,,1,,-1,1,,2 \\ 
drayage-25-23 & 0.23 & 0.43 & ,,,3,0,,-1,,,0,,2,2,3,2,2,0,-1,4,,,-1,,2,-1,,2,,3,,,,3,8,,0,3,-1,-1,1,,,0,5,-1,,2 \\ 
n3div36 & 12.43 & 18.89 & ,,,1,,,-1,,1,,,-1,-1,-1,-1,-1,0,1,-1,,-1,,,1,-1,,,,-1,,,,,8,,,0,-1,1,1,,1,,-1,-1,,0 \\ 
ns2996139 & 17.87 & 15.95 & ,,,,,,,,,,,,,,,,,,,,,,,2,,,,,,,,,,8,,,,,,,,,,,,, \\ 
ns3360030 & 82.55 & 156.53 & ,,,,,,,,,,,,,,,,,,,,,,,1,,,,,,,,,,8,,,,,,,,,,,,, \\ 
net12 & 14.35 & 36.21 & ,,,,0,,,,,,,1,1,-1,-1,,,-1,,,,1,,1,,,,,-1,,,,,8,,,,,-1,,,1,,-1,-1,, \\ 
neos-5195221-niemur & 242.48 & 1845.94 & ,,,,2,,1,,2,,,-1,2,-1,-1,1,4,1,4,,,1,,2,2,,,,-1,,,,2,8,,0,2,2,-1,2,,1,,-1,2,,0 \\ 
neos-3381206-awhea & 0.29 & 1.91 & ,,,,2,,-1,,,1,,3,-1,-1,2,2,4,2,4,,1,-1,,2,-1,,,,-1,,,,3,8,,1,0,,-1,-1,,1,0,5,1,,2 \\ 
eilA101-2 & 303.15 & 3029.9 & ,,,3,,,,,,1,,-1,1,1,1,-1,0,-1,-1,,1,,,2,2,,5,,-1,,0,,3,8,,,3,,1,,,4,,-1,-1,, \\ 
ns2441809 & 102.9 & 174.54 & ,,,1,,,1,,,,,2,2,,,-1,,,,,2,,,2,,,,,-1,,,,,10,,,0,-1,-1,-1,,1,0,,-1,, \\ 
leo1 & 72.08 & 158.95 & ,,,,,,,,,,,,-1,-1,,-1,,1,,,2,,,2,2,,1,,-1,,,,,8,,,,1,,2,,,0,,,, \\ 
ex10 & 8.47 & 8.37 & ,,,3,,,3,,2,1,,2,-1,1,-1,-1,5,2,,,-1,-1,,0,2,,6,,3,,,,,8,,0,0,1,-1,-1,,,,5,2,,2 \\ 
neos-1122047 & 0.42 & 1.73 & ,,,1,2,,-1,,1,,,-1,1,-1,,1,1,1,1,,2,,,0,1,,1,,3,,1,,1,8,,1,3,1,1,1,,3,,4,-1,,2 \\ 
ns4228793 & 5590 & 10800.16 & ,,,,0,,-1,,1,1,,2,2,1,-1,2,3,1,4,,2,-1,,2,2,,2,,-1,,,,0,4,,,0,,-1,1,,3,,-1,2,,2 \\ 
sing326 & 440.03 & 1659.62 & ,,,,0,,-1,,,,,-1,,-1,-1,,,1,,,-1,-1,,1,1,,,,1,,,,,8,,,0,,1,1,,,,,1,,0 \\ 
graph20-20-1rand & 17.75 & 68.63 & ,,,,,,-1,,,1,,3,-1,1,2,2,,2,3,,1,-1,,1,1,,6,,3,,,,,8,,,3,2,1,-1,,1,,-1,,,2 \\ 
stein45inf & 0.03 & 0.07 & ,,,1,3,,-1,,,,,2,-1,1,-1,-1,,-1,4,,-1,1,,2,-1,,,,3,,0,,,8,,1,3,2,1,2,,4,,5,2,,2 \\ 
radiationm18-12-05 & 268.43 & 486.37 & ,,,,,,3,,,1,,-1,1,2,-1,2,,2,4,,2,,,2,2,,6,,3,,,,,8,,1,3,-1,-1,2,,,,5,2,,0 \\ 
neos859080 & 0.03 & 0.11 & ,,,2,1,,1,,2,0,,,1,1,-1,-1,2,,1,,,-1,,0,-1,,3,,1,,,,2,8,,0,,,1,-1,,3,0,3,1,,2 \\ 
neos-5114902-kasavu & 937.13 & 7200 & ,,,,3,,-1,,,0,-1,-1,2,3,2,-1,5,2,,3,2,-1,-1,0,-1,,6,,-1,2,0,,3,8,5,1,3,1,,2,3,4,0,5,-1,,0 \\ 
ns2164569 & 553.99 & 1037.26 & ,,,,,,2,,,1,,1,1,1,-1,,0,,,,,1,,2,-1,,,,1,,,,0,8,,,,,1,-1,,1,0,-1,,,0 \\ 
triptim1 & 10.09 & 31.06 & ,,,,0,,-1,,,1,,-1,-1,-1,,-1,0,-1,2,,-1,-1,,0,-1,,3,,3,,,,0,8,,1,,2,,2,,1,0,5,,, \\ 
hypothyroid-k1 & 2.07 & 21.75 & ,,,1,1,,2,,2,,,1,1,,,1,1,-1,3,,,1,,0,,,3,,,,2,,0,5,,,3,,-1,-1,,1,0,3,1,,2 \\ 
k1mushroom & 59.73 & 509.46 & ,,,3,3,,3,,,,,2,1,-1,2,,1,2,-1,,1,,,2,-1,,,,-1,,0,,,8,,1,,-1,,,,4,,-1,1,,2 \\ 
markshare\_4\_0 & 0.2 & 2.58 & ,,,3,,,-1,,,0,,-1,-1,3,-1,1,,2,-1,,-1,1,,0,-1,,6,,1,,,,3,8,,0,0,-1,-1,2,,1,0,-1,2,, \\ 
chromaticindex1024-7 & 91.84 & 468.07 & ,,,,,,,,,,,,,,-1,,,,,,,,,2,,,,,,,,,,8,,,0,,,,,1,0,,,,2 \\ 
fhnw-binpack4-4 & 1423.93 & 5667.75 & ,,,,1,,-1,,,1,,1,-1,1,,-1,,,,,-1,,,2,2,,1,,3,,,,,8,,1,0,2,1,2,,2,0,1,,,0 \\ 
tr12-30 & 20.05 & 29.21 & ,,,,,,,,,,,,,,-1,,,-1,,,,,,2,,,,,,,,,,8,,0,,,,,,,0,-1,1,, \\ 
neos-860300 & 1.8 & 3.74 & ,,,2,1,,1,,2,0,,1,,1,,,2,,2,,,,,1,,,3,,1,,2,,1,8,,0,2,,,,,2,0,2,-1,, \\ 
roll3000 & 3.34 & 6.55 & ,,,,,,-1,,,,,,-1,-1,,1,,1,,,-1,-1,,2,,,,,1,,,,,8,,,0,1,1,,,,,-1,1,,0 \\ 
neos-4738912-atrato & 3.46 & 19.29 & ,,,2,0,,3,,,1,,2,1,2,2,,5,2,,,-1,1,,0,,,1,,-1,,0,,3,8,,,,,-1,,,2,,,1,,2 \\ 
lotsize & 48.46 & 93.9 & ,,,,,,,,,,,,,,,,,,,,,,,2,1,,,,,,,,,8,,,,,,,,,,,,, \\ 
nexp-150-20-8-5 & 15.83 & 159.03 & ,,,2,3,,2,,2,0,,,-1,-1,1,1,,1,,,,1,,0,,,4,,3,,1,,,8,,0,3,2,1,1,,2,0,,1,, \\ 
fhnw-binpack4-18 & 0.27 & 0.45 & ,,,,1,,,,,1,,-1,2,,2,,5,-1,-1,,2,-1,,2,,,,,-1,,,,,7,,,,2,-1,-1,,4,,5,-1,,0 \\ 
u30t24ramp & 31.55 & 34.17 & ,,,,1,,,,,0,,,,,,,,,,,,,,2,,,3,,,,,,,8,,,,,,,,,,,,, \\ 
ns2082847 & 0.01 & 0.06 & ,,,3,3,,-1,,1,1,,2,,2,1,,3,2,2,,-1,-1,,2,,,3,,2,,,,0,2,,0,0,-1,1,1,,2,0,3,,, \\ 
roi5alpha10n8 & 513.91 & 1075.09 & ,,,,3,,3,,,,,3,-1,3,,-1,5,2,4,,2,,,2,,,,,,,,,3,8,,,,1,1,1,,,0,-1,-1,,2 \\ 
mushroom-best & 45.47 & 93.21 & ,,,,1,,-1,,,,,,,-1,,-1,,-1,,,1,-1,,1,,,,,-1,,,,,8,,,,,,,,,,-1,,,0 \\ 
maritime-jg3d9 & 0.4 & 0.69 & ,,,,3,,,,4,,,1,2,-1,2,-1,1,,4,,1,,,0,-1,,,,3,,,,0,3,,1,2,,1,-1,,,0,4,-1,, \\ 
neos-4300652-rahue & 118.73 & 2063.06 & ,,,1,3,,2,,,,,3,2,3,,1,4,1,4,,1,1,,1,,,,,-1,,,,1,7,,0,0,1,,-1,,1,,-1,-1,,0 \\ 
sp97ar & 354.56 & 416.52 & ,,,,,,,,,,,,,,,,,,,,,1,,2,,,,,,,,,,8,,,,,,,,,,,,, \\ 
highschool1-aigio & 517.5 & 7200 & ,,,,,,-1,,,,,3,1,-1,2,2,,-1,-1,,2,-1,,1,1,,6,,3,,,,3,8,,1,0,2,-1,2,,,,5,1,, \\ 
u50t24wc & 192.98 & 63.96 & ,,,,,,,,,,,,,,,,,,,,,,,2,,,,,,,,,,8,,,,,,,,,,,,, \\ 
supportcase7 & 5.02 & 13.85 & ,,,,,,-1,,,,,3,2,-1,-1,2,5,,4,,-1,,,0,-1,,2,,,,,,,8,,1,2,2,-1,-1,,4,0,-1,-1,, \\ 
rmatr100-p10 & 1.88 & 3.06 & ,,,3,,,-1,,,1,,-1,-1,2,2,-1,,2,4,,2,-1,,2,2,,,,-1,,0,,3,8,,1,3,-1,-1,2,,1,0,-1,2,,0 \\ 
brazil3 & 32.58 & 544.67 & ,,,,,,1,,1,0,,,,,,,,,-1,,,-1,,1,-1,,,,1,,,,,8,,,0,-1,,1,,,,-1,-1,,2 \\ 
supportcase29 & 1.86 & 21.3 & ,,,3,3,,,,1,0,,3,,,2,2,0,,4,,,1,,2,,,,,3,,,,2,8,,1,,2,1,1,,4,0,,,,2 \\ 
seymour & 1456.89 & 2067.45 & ,,,,,,,,,,,,,,,,,,,,,,,2,,,,,,,,,,8,,,,1,,,,,,,,, \\ 
proteindesign121hz512p9 & 422.17 & 3083.31 & ,,,,0,,-1,,,,,1,2,,1,-1,0,,,,-1,,,0,2,,1,,2,,,,3,8,,1,3,2,1,,,,,-1,2,, \\ 
ns1208400 & 20.21 & 27.81 & ,,,3,,,-1,,,,,-1,-1,-1,2,2,5,1,4,,2,,,1,-1,,6,,3,,,,3,8,,,0,1,,-1,,3,,-1,-1,,2 \\ 
netdiversion & 21.83 & 51.39 & ,,,,2,,2,,,0,,1,,-1,2,-1,0,,3,,2,,,1,2,,6,,,,,,3,8,,,,,-1,,,,,1,-1,, \\ 
ns4636843 & 44.54 & 321.4 & ,,,,,,,,,,,-1,1,,,,,1,,,-1,1,,1,-1,,,,1,,,,,8,,,0,,,,,,,-1,1,,0 \\ 
satellites2-60-fs & 142.17 & 563.56 & ,,,,0,,1,,,0,,,1,,-1,-1,,,,,,,,2,1,,,,,,,,,8,,,0,1,,-1,,1,,,,, \\ 
neos-1456979 & 15.04 & 35.97 & ,,,,0,,-1,,,1,,-1,,,,-1,,-1,,,-1,,,1,,,1,,-1,,,,,8,,,,1,1,,,,,,,, \\ 
physiciansched3-3 & 3780.16 & 5616.19 & ,,,,,,,,,,,,,,,-1,,,,,,,,2,,,,,,,,,,8,,,,,,,,,,,,, \\ 
app1-1 & 0.11 & 0.16 & ,,,3,3,,,,,,,-1,-1,3,1,-1,5,-1,4,,2,1,,2,2,,,,,,,,0,8,,,0,2,1,2,,,,-1,2,,2 \\ 
ns2394796 & 0.8 & 4.04 & ,,,1,3,,-1,,,,,3,-1,-1,2,1,0,2,4,,2,,,0,-1,,,,3,,,,3,12,,1,0,2,,-1,,,,-1,2,,2 \\ 
ns2494475 & 0.66 & 1.49 & ,,,,,,,,,,,,,,,,,,,,,,,2,,,,,,,,,,8,,,,,,,,,,,,, \\ 
supportcase12 & 1097.48 & 1814.58 & ,,,3,3,,3,,,1,,2,2,3,1,2,,2,-1,,2,,,2,,,,,,,,,2,8,,1,0,2,-1,-1,,4,0,-1,1,,0 \\ 
co-100 & 80.96 & 85.75 & ,,,,,,-1,,,0,,-1,-1,,1,,,,,,,-1,,2,1,,,,,,,,,8,,,2,1,1,,,,0,-1,-1,, \\ 
u40t24ramp & 263.23 & 366.85 & ,,,,,,,,,,,,,,,,,,,,,,,2,,,,,-1,,,,,8,,,,,,,,,,,,, \\ 
neos-662469 & 23.4 & 40.97 & ,,,,0,,-1,,,,,,,,,-1,,-1,,,,,,2,,,,,-1,,,,,8,,,,-1,,,,,,-1,1,, \\ 
neos-4532248-waihi & 856.91 & 7200 & ,,,,0,,1,,1,0,,-1,-1,3,1,,,-1,,,2,-1,,0,1,,,,2,,,,1,8,,,0,-1,,1,,1,0,-1,,, \\ 
bnatt400 & 458.31 & 741.9 & ,,,,,,,,,,,,,,,,,,,,,,,2,,,,,,,,,,8,,,,,,,,,,,,, \\ 
graphdraw-domain & 24.77 & 19.28 & ,,,3,2,,3,,,1,,-1,-1,3,2,,5,1,,,1,-1,,1,2,,2,,1,,2,,,8,,1,3,2,,,,1,0,,2,,0 \\ 
cryptanalysiskb128n5obj14 & 821.07 & 7200 & ,,,1,0,,1,,1,1,,-1,-1,2,2,1,4,,4,,2,1,,1,,,4,,1,,1,,2,8,,0,,2,,2,,1,,3,-1,,0 \\ 
gfd-schedulen180f7d50m30k18 & 102.65 & 7200 & ,,,,3,,2,,1,0,,1,2,1,,,,,2,,,,,0,2,,,,2,,2,,1,8,,1,,,,2,,3,,4,,, \\ 
leo2 & 135.02 & 250.24 & ,,,,,,,,,,,,,,,,0,,,,,,,2,,,,,1,,,,0,8,,,,,,,,,,,,, \\ 
rail507 & 6.27 & 18.78 & ,,,,,,,,,,,-1,,,-1,,1,,,,,,,1,,,,,2,,,,0,8,,,,1,,,,,0,-1,-1,, \\ 
proteindesign122trx11p8 & 363.4 & 967.7 & ,,,,1,,-1,,,,,,-1,,2,2,,-1,,,-1,-1,,2,-1,,,,,,,,,8,,,,1,,,,,,-1,,,0 \\ 
milo-v12-6-r2-40-1 & 123.32 & 3764.53 & ,,,,,,,,1,1,,3,,,2,,,,,,-1,1,,2,,,,,,,,,0,8,,0,3,2,1,1,,,,-1,-1,,0 \\ 
neos-1582420 & 0.72 & 2.37 & ,,,1,,,-1,,,,,2,2,3,-1,-1,4,2,4,,2,-1,,2,,,,,,,,,0,8,,,,2,-1,2,,,,,,, \\ 
stein9inf & 0 & 0.01 & ,,,1,2,,3,,4,,,2,2,3,-1,,4,-1,-1,,2,1,,1,,,6,,,,1,,1,4,,,0,2,-1,,,2,0,-1,1,, \\ 
dell & 0 & 0 & ,,,,2,,-1,,4,1,,3,1,-1,2,1,1,1,2,,2,1,,0,-1,,4,,2,,1,,1,1,,,2,,1,2,,4,0,1,1,, \\ 
neos-5049753-cuanza & 663.84 & 925.17 & ,,,3,3,,2,,,0,,3,-1,-1,1,2,4,2,,,,-1,,0,-1,,6,,3,,1,,3,8,,1,2,-1,1,2,,,0,-1,1,, \\ 
CMS750\_4 & 4.72 & 17.67 & ,,,1,,,1,,,0,,3,,-1,,2,,-1,,,1,,,2,-1,,,,3,,,,,8,,1,0,2,-1,1,,,0,-1,-1,,0 \\ 
enlight4 & 0 & 0 & ,,,1,,,,,3,1,,2,,-1,-1,,4,,3,,1,1,,2,1,,4,,,,2,,0,6,,0,3,1,-1,-1,,,0,4,-1,,2 \\ 
supportcase33 & 6.38 & 28.2 & ,,,,,,1,,,0,,1,,-1,,1,,1,,,1,-1,,1,,,1,,,,,,,8,,,3,1,1,2,,1,,,2,,2 \\ 
qap10 & 1.06 & 3.26 & ,,,2,,,1,,4,0,,,2,-1,-1,-1,2,-1,2,,1,,,0,,,4,,3,,1,,0,8,,,,-1,1,1,,,0,5,,,0 \\ 
irp & 0.53 & 1.29 & ,,,1,3,,1,,,0,,-1,1,3,1,1,2,,4,,1,-1,,1,,,,,-1,,1,,0,8,,1,2,2,,,,1,,3,2,,0 \\ 
istanbul-no-cutoff & 6.92 & 16.42 & ,,,1,,,,,,,,3,,1,1,2,,,,,1,,,2,1,,1,,-1,,,,,8,,0,0,,-1,-1,,1,,,1,,2 \\ 
ns2382816 & 5.97 & 721.94 & ,,,,0,,,,,,,1,-1,2,2,2,,,-1,,-1,-1,,0,-1,,,,2,,,,,8,,0,2,2,1,,,1,0,-1,-1,,2 \\ 
30n20b8 & 0.97 & 1.66 & ,,,,,,,,,0,,,1,-1,1,1,0,1,,,2,-1,,0,,,,,-1,,,,,8,,,,-1,,-1,,,,-1,-1,,2 \\ 
stein15inf & 0.01 & 0.02 & ,,,,,,,,,0,,3,1,,,1,2,1,4,,1,1,,0,-1,,6,,,,0,,1,8,,0,,-1,,2,,2,0,-1,1,, \\ 
neos-827175 & 0.46 & 1.36 & ,,,3,3,,2,,2,,,-1,2,2,-1,,0,1,1,,-1,-1,,1,,,1,,3,,0,,2,8,,1,0,-1,,,,,0,4,-1,, \\ 
ns2369235 & 25.33 & 32.51 & ,,,,0,,-1,,1,,,,,,,,,,,,,,,2,1,,,,1,,,,,8,,,2,,1,-1,,,,,1,, \\ 
rail01 & 132.19 & 502.61 & ,,,,,,,,,,,,,,,,,,,,,,,2,,,,,1,,,,,8,,,,,,,,,,,,, \\ 
radiationm40-10-02 & 1032.17 & 7200 & ,,,1,,,,,,1,,,1,2,,1,5,2,,,,,,2,,,3,,,,,,3,6,,1,3,,1,1,,4,0,5,1,,0 \\ 
unitcal\_7 & 17.97 & 39.42 & ,,,1,3,,-1,,,,,3,1,3,2,-1,,1,3,,1,,,2,,,,,3,,1,,,8,,,0,2,-1,1,,,,-1,-1,,2 \\ 
neos-2978193-inde & 7.83 & 17.07 & ,,,,,,,,,,,-1,1,1,-1,,,-1,,,-1,,,2,,,,,,,,,,8,,0,0,-1,-1,-1,,,,,1,,0 \\ 
fiball & 1.74 & 10.88 & ,,,2,3,,1,,,,,1,1,,2,1,3,,3,,,-1,,0,-1,,5,,,,1,,2,8,,1,2,,,-1,,1,0,3,-1,, \\ 
u40t24wc & 23.8 & 28.46 & ,,,3,3,,,,,,,2,-1,2,-1,2,4,2,4,,-1,1,,1,-1,,,,-1,,,,,12,,1,3,2,-1,-1,,,,-1,-1,,2 \\ 
bc & 20.47 & 41.78 & ,,,3,3,,-1,,,,,3,2,3,2,2,5,2,4,,1,-1,,1,,,,,3,,,,3,1,,1,0,2,-1,-1,,,,-1,1,,2 \\ 
mcsched & 3.08 & 14.72 & ,,,,,,,,,,,-1,,1,,,,,,,-1,-1,,2,,,,,,,,,0,8,,,0,-1,,-1,,,,,-1,,0 \\ 
g503inf & 0 & 0 & ,,,1,1,,,,1,1,,3,1,3,1,,,,-1,,,-1,,0,-1,,2,,3,,,,1,3,,,3,-1,,1,,4,,1,2,,2 \\ 
uccase12 & 7.48 & 16.37 & ,,,2,,,,,,,,2,-1,-1,-1,-1,,1,-1,,-1,-1,,0,-1,,,,3,,,,,8,,,2,-1,-1,-1,,1,,-1,-1,,0 \\ 
ns3555904 & 138.55 & 10800.21 & ,,,3,3,,1,,,0,,1,1,1,,,1,,1,,2,1,,1,,,1,,,,,,3,8,,1,,,,1,,2,,2,-1,,0 \\ 
pk1 & 1.68 & 6.2 & ,,,,,,-1,,,1,,3,2,-1,2,2,,-1,,,2,-1,,2,1,,,,3,,,,2,8,,,3,-1,-1,2,,4,,-1,2,,2 \\ 
neos-2987310-joes & 0.84 & 2.57 & ,,,1,3,,-1,,,1,,-1,-1,-1,-1,-1,,-1,3,,2,-1,,2,-1,,,,3,,0,,,8,,,3,-1,-1,2,,,0,1,-1,,2 \\ 
neos-3402294-bobin & 7.13 & 101 & ,,,1,,,,,,,,-1,1,1,1,-1,,2,1,,1,1,,0,-1,,5,,-1,,,,,1,,0,3,-1,1,2,,4,,5,-1,,0 \\ 
eil33-2 & 1.76 & 9.6 & ,,,1,3,,,,1,0,,-1,-1,-1,1,,2,1,,,,1,,1,1,,5,,,,,,2,8,,0,0,,-1,,,3,,4,,, \\ 
buildingenergy & 207.96 & 1216.14 & ,,,1,1,,-1,,,,,,-1,,,-1,,-1,,,-1,,,1,1,,3,,,,2,,,6,,0,2,1,1,1,,,,-1,1,,0 \\ 
beasleyC3 & 2.12 & 8.52 & ,,,,,,-1,,,1,,-1,,3,-1,-1,0,-1,-1,,-1,1,,2,-1,,1,,1,,,,,8,,,2,1,1,1,,,,-1,,,2 \\ 
thor50dday & 276.43 & 3734.75 & ,,,3,2,,,,2,,,,1,1,-1,-1,4,-1,1,,1,-1,,1,1,,,,-1,,1,,1,8,,1,,-1,1,-1,,2,0,-1,2,,0 \\ 
swath3 & 1.75 & 3.72 & ,,,2,,,2,,,,,-1,1,,-1,1,,1,,,,1,,1,-1,,,,3,,,,,7,,,3,2,1,-1,,,,-1,,, \\ 
rd-rplusc-21 & 73.98 & 179.49 & ,,,,3,,-1,,,1,,3,2,-1,2,2,,,-1,,2,,,0,-1,,3,,3,,,,,8,,,3,2,-1,-1,,3,,-1,2,, \\ 
neos-2746589-doon & 19.21 & 48.21 & ,,,,1,,-1,,,1,,1,1,,,-1,,1,,,1,-1,,0,,,2,,,,,,0,8,,,,,1,-1,,,,,2,,0 \\ 
neos-3083819-nubu & 0.37 & 1.54 & ,,,,,,-1,,,1,,2,-1,-1,-1,2,,2,,,2,1,,0,-1,,2,,-1,,,,2,8,,,,2,-1,2,,4,0,-1,2,,0 \\ 
ns2070961 & 269.17 & 548.33 & ,,,,,,-1,,,,,-1,-1,,,1,0,,,,-1,,,1,,,,,-1,,,,1,7,,0,,,-1,2,,2,,-1,-1,,0 \\ 
var-smallemery-m6j6 & 159.79 & 133.63 & ,,,3,,,-1,,,0,,,-1,,1,-1,5,-1,2,,,,,2,-1,,2,,1,,1,,0,8,,1,3,-1,1,1,,4,,-1,,, \\ 
map16715-04 & 79.39 & 60.13 & ,,,,0,,,,,,,,,,1,,,1,,,-1,1,,2,-1,,,,-1,,,,,8,,0,0,1,,-1,,,0,-1,,,2 \\ 
germanrr & 170.98 & 213.07 & ,,,,0,,,,,0,,,1,,,1,,,,,1,,,2,,,,,,,,,,8,,0,0,1,,,,,0,,,,2 \\ 
mc11 & 2.38 & 12.03 & ,,,,,,-1,,,1,,1,2,,-1,2,4,-1,-1,,-1,,,0,1,,2,,-1,,,,0,8,,1,3,2,-1,-1,,4,,-1,2,,0 \\ 
academictimetablesmall & 361.18 & 682.32 & ,,,,,,,,,,,,,,,,,,,,,-1,,2,,,,,-1,,,,,8,,,2,1,,,,,,,,,0 \\ 
assign1-5-8 & 43.28 & 1364.34 & ,,,,0,,,,,,,-1,,1,,,,-1,,,1,-1,,2,,,,,1,,,,,8,,0,0,1,,-1,,,,,2,,0 \\ 
cmflsp50-24-8-8 & 855.33 & 2591.67 & ,,,,,,1,,2,0,,,,,,,,-1,,,,,,2,1,,,,,,,,,8,,0,2,-1,,,,,0,,-1,, \\ 
enlight\_hard & 0.01 & 0.01 & ,,,,,,,,,,,,,,,,,,,,,,,2,,,,,,,,,,8,,,,,,,,,,,,, \\ 
oocsp-racks030f7cci & 1.85 & 3.88 & ,,,2,,,2,,,,,3,-1,,-1,,5,2,3,,2,,,0,,,6,,-1,,,,2,1,,1,0,-1,-1,2,,,0,-1,-1,,0 \\ 
neos5 & 1.74 & 5.84 & ,,,,,,-1,,,0,,,1,-1,2,2,,1,,,,1,,1,-1,,,,3,,,,,8,,,0,,1,2,,,0,-1,2,,2 \\ 
bppc4-08 & 49.51 & 2556.65 & ,,,,3,,,,,1,,-1,-1,3,-1,,5,1,,,2,,,0,-1,,1,,2,,,,,8,,0,0,,1,1,,1,0,-1,1,,2 \\ 
ns2071214 & 100.19 & 173.92 & ,,,2,,,-1,,,,,1,,,-1,1,,1,4,,1,,,2,2,,,,3,,1,,0,12,,,3,2,,-1,,1,,,2,,2 \\ 
ns1760995 & 117.85 & 249.3 & ,,,,,,-1,,,,,1,,1,,1,,1,-1,,-1,1,,2,1,,,,1,,,,,8,,,2,1,,-1,,,,-1,1,,2 \\ 
sing44 & 776.89 & 4882.4 & ,,,,0,,-1,,2,0,,2,,,2,,5,2,,,-1,-1,,2,2,,5,,,,0,,3,8,,1,3,2,-1,1,,4,0,,2,,2 \\ 
s250r10 & 46.25 & 2546.63 & ,,,3,2,,-1,,,,,-1,,-1,1,1,4,1,3,,2,,,1,1,,3,,,,1,,2,8,,,3,,1,2,,2,,3,-1,, \\ 
neos-859770 & 0.94 & 2.64 & ,,,,,,,,,,,1,,-1,,1,,1,-1,,1,-1,,2,-1,,,,-1,,,,,8,,0,,,1,-1,,1,,-1,1,,2 \\ 
nw04 & 1 & 2.6 & ,,,,3,,1,,,0,,-1,2,3,2,-1,0,2,4,,-1,-1,,2,-1,,6,,-1,,,,3,8,,0,,2,-1,1,,,0,-1,2,,2 \\ 
app1-2 & 6.7 & 10.49 & ,,,,,,1,,,,,1,,-1,,1,,-1,,,2,1,,2,-1,,,,1,,,,,8,,,2,,-1,-1,,,0,,-1,, \\ 
n5-3 & 0.47 & 1.46 & ,,,2,,,1,,,,,3,1,,-1,,,,,,1,1,,0,,,1,,1,,,,,8,,,,,-1,-1,,,,,,,0 \\ 
csched008 & 30.54 & 186.43 & ,,,3,0,,,,,0,,2,-1,,2,,4,-1,2,,-1,,,0,-1,,6,,1,,,,1,8,,1,,2,1,,,2,,,,, \\ 
mod008inf & 0.01 & 0.02 & ,,,,3,,1,,,0,,,2,2,-1,,5,2,-1,,-1,-1,,2,,,6,,1,,,,3,3,,0,3,2,1,,,4,0,2,-1,, \\ 
ns2326618 & 279.03 & 810.75 & ,,,,0,,,,,0,,-1,-1,-1,,2,0,-1,,,-1,,,2,,,,,,,,,,8,,,3,,-1,,,,0,-1,1,, \\ 
square41 & 131.83 & 201.71 & ,,,,,,,,,0,,,,,-1,-1,,,,,-1,,,2,1,,,,-1,,,,0,8,,0,,,1,1,,1,,-1,1,, \\ 
rococoC10-001000 & 12.5 & 15.43 & ,,,,,,,,,,,,1,-1,,,,,,,,,,2,,,,,-1,,,,0,8,,,,-1,,,,,,,-1,, \\ 
neos-5157194-moruya & 1.67 & 2.65 & ,,,2,,,3,,,1,,3,2,-1,-1,2,5,2,-1,,2,1,,2,2,,,,3,,1,,0,8,,,3,-1,-1,1,,,,-1,2,,0 \\ 
uct-subprob & 224.22 & 298.22 & ,,,,,,,,,,,,,,,,,,,,,,,2,,,,,,,,,,8,,,,,,,,,,,,, \\ 
fastxgemm-n2r6s0t2 & 0.81 & 22.61 & ,,,,3,,1,,,1,,3,,3,2,2,5,-1,,,1,1,,0,2,,3,,-1,,,,0,8,,,0,2,,-1,,2,0,-1,-1,,0 \\ 
dano3\_5 & 39.29 & 73.43 & ,,,3,0,,-1,,,0,,1,1,,2,-1,,1,-1,,-1,1,,1,-1,,5,,,,,,3,8,,,3,2,,2,,4,0,-1,2,,2 \\ 
opm2-z10-s4 & 1554.63 & 3276.97 & ,,,3,3,,-1,,,,,2,,3,-1,2,5,2,4,,2,1,,0,-1,,,,3,,,,,8,,0,0,2,-1,-1,,1,,-1,-1,,0 \\ 
b1c1s1 & 604.45 & 1244.58 & ,,,,,,,,,,,-1,,,,,,1,,,,-1,,2,,,,,1,,,,,8,,,2,,1,,,,0,,,,2 \\ 
ns2081729 & 7.47 & 33.09 & ,,,,,,,,,,,,1,,,,,,,,,,,1,,,,,,,,,,8,,,,,,,,,,,,, \\ 
binkar10\_1 & 0.38 & 0.7 & ,,,3,,,1,,,1,,2,,1,-1,-1,,2,4,,-1,1,,1,2,,6,,3,,,,,8,,,2,2,-1,1,,,,-1,1,,2 \\ 
exp-1-500-5-5 & 0.34 & 1.23 & ,,,,,,1,,,,,,1,1,1,,,,,,,-1,,2,1,,,,1,,,,,8,,0,0,,,-1,,1,0,,2,,0 \\ 
swath1 & 0.31 & 0.68 & ,,,,,,1,,,,,,1,2,-1,1,0,-1,,,2,-1,,2,-1,,,,,,,,,8,,1,3,1,-1,1,,,,-1,-1,,0 \\ 
neos-3555904-turama & 69.25 & 7200 & ,,,3,3,,1,,,0,,1,1,1,,,1,,1,,2,1,,1,,,1,,,,,,3,8,,1,,,,1,,2,,2,-1,,0 \\ 
chrom\_1024 & 102.88 & 206.28 & ,,,,,,-1,,,,,1,-1,-1,1,,,1,-1,,-1,1,,2,-1,,,,-1,,,,0,9,,,2,1,1,1,,,0,-1,,, \\ 
neos-4338804-snowy & 219.63 & 358.37 & ,,,1,0,,1,,,,,,,,,,0,-1,,,,,,2,,,1,,,,,,,8,,,2,,,,,1,,,,, \\ 
neos-3216931-puriri & 5.54 & 230.11 & ,,,,2,,1,,2,,,,-1,,-1,1,0,1,3,,1,-1,,2,2,,4,,-1,,1,,2,7,,0,2,1,-1,2,,,,3,2,, \\ 
mas76 & 1.29 & 6.76 & ,,,,1,,1,,,,,2,,-1,2,,0,-1,,,1,1,,2,2,,1,,-1,,2,,0,8,,1,0,-1,-1,1,,,,-1,1,,2 \\ 
traininstance2 & 182.29 & 666.61 & ,,,,0,,-1,,,,,,,-1,-1,1,,,,,1,1,,2,,,,,,,,,0,8,,0,0,1,,-1,,,,,-1,, \\ 
neos-4382714-ruvuma & 3600 & 3600.02 & ,,,,,,,,,,,,,,,,,,,,,,,,,,,,,,,,,,,,,,,,,,,,,, \\ 
ns2319318 & 183.7 & 1148.35 & ,,,,3,,-1,,,0,,3,1,2,-1,2,,2,4,,2,1,,0,-1,,1,,-1,,0,,,4,,,0,,,2,,4,0,5,-1,,2 \\ 
cvs16r128-89 & 482.34 & 596.24 & ,,,1,0,,-1,,,0,,,1,,,-1,,-1,,,,-1,,2,,,1,,-1,,,,0,7,,0,,,-1,-1,,,0,,,, \\ 
splice1k1 & 507.67 & 7200 & ,,,3,,,1,,2,1,,-1,-1,1,,2,,1,4,,1,,,1,,,6,,1,,1,,0,8,,0,0,,-1,2,,,,-1,-1,,0 \\ 
uc720-7-4-4-8 & 88.97 & 3694.75 & ,,,3,2,,-1,,5,,,-1,2,-1,2,-1,,,-1,,-1,1,,0,1,,1,,-1,,1,,,1,,0,,2,-1,-1,,,,-1,2,, \\ 
neos-4387871-tavua & 1741.8 & 4009.87 & ,,,,0,,,,,,,-1,,,,-1,,,,,-1,-1,,2,,,1,,3,,,,,8,,,3,2,,-1,,,,,,, \\ 
ns476799 & 104.7 & 726.51 & ,,,,1,,-1,,1,0,,-1,2,-1,1,1,,-1,2,,2,,,0,1,,,,3,,,,1,12,,,3,-1,1,1,,1,,,2,, \\ 
neos-3004026-krka & 4.04 & 76.92 & ,,,,1,,-1,,5,1,,3,2,-1,2,2,,-1,4,,-1,1,,1,2,,6,,2,,,,3,8,,,0,2,-1,-1,,4,0,5,2,,0 \\ 
50v-10 & 158.3 & 243.5 & ,,,,,,-1,,,,,1,1,,,,,1,,,-1,-1,,2,,,,,1,,,,,8,,,,1,,,,,,-1,1,, \\ 
ns2267839 & 730.72 & 562.38 & ,,,,,,1,,,,,1,,-1,-1,-1,,,,,,,,2,1,,,,1,,,,,9,,,2,-1,,-1,,,,-1,-1,,2 \\ 
supportcase19 & 509.39 & 952.47 & ,,,,2,,-1,,,0,,-1,-1,-1,,-1,,,1,,2,,,1,2,,6,,-1,,,,3,8,,,0,,,2,,4,,5,2,, \\ 
ran14x18-disj-8 & 25.01 & 56.99 & ,,,2,,,,,,0,,1,-1,,,-1,0,,,,-1,,,2,1,,1,,,,1,,,8,,,,1,,1,,,0,,,,2 \\ 
air05 & 1.28 & 2.59 & ,,,,,,-1,,,,,,-1,-1,,,,1,,,1,,,1,-1,,,,1,,,,,8,,0,,1,1,,,,,-1,2,, \\ 
neos-4763324-toguru & 5809.21 & 7200 & 1,2,1,3,3,100,3,5,2,,0,,,,,,,,4,,1,-1,1,1,2,4,2,2,,5,,5,1,7,4,1,0,-1,1,2,-1,4,0,,,0,2 \\ 
neos-2626858-aoos & 0.5 & 0.82 & ,,,3,,,-1,,,1,,-1,2,,1,2,,2,4,,2,1,,2,-1,,1,,3,,,,2,8,,,,2,1,,,4,,5,1,,0 \\ 
ns1943024 & 265.11 & 430.89 & ,,,,,,-1,,,,,,-1,,,,,,,,-1,,,2,,,,,-1,,,,,8,,,,,,,,,0,-1,,,0 \\ 
gen-ip002 & 26.79 & 187.25 & ,,,1,0,,,,2,1,,,,-1,,-1,,-1,,,,1,,0,,,1,,,,,,0,8,,0,,-1,,1,,,0,,1,, \\ 
ns1952667 & 2.42 & 78.21 & ,,,2,3,,-1,,1,0,,-1,,-1,2,-1,1,2,4,,-1,1,,2,-1,,5,,-1,,0,,,8,,1,0,-1,1,2,,,0,5,2,,2 \\ 
misc04inf & 0.03 & 0.12 & ,,,3,,,-1,,,,,,2,3,,-1,,2,1,,2,,,2,,,1,,-1,,1,,3,1,,0,0,2,,-1,,,,,-1,,0 \\ 
neos-1445765 & 0.86 & 1.74 & ,,,3,0,,3,,,,,-1,2,-1,2,2,1,2,4,,2,,,0,-1,,5,,3,,1,,3,8,,,2,2,1,2,,,,2,-1,,0 \\

\end{longtable}

\section{Related Work}
Automated configuration of MILP solvers has attracted significant attention in recent years. This field has evolved rapidly, with various teams contributing methods and insights to enhance solver performance. Paper~\cite{hutter2010automated} laid a foundation of automated configuration. Building on this, Hydra-MIP~\cite{xu2011hydra} introduced an innovative machine-learning approach to automated algorithm configuration for MIP. SMAC3~\cite{lindauer2022smac3} offers a versatile software package for fine-tuning arbitrary algorithms through Bayesian optimization techniques. MindOpt Tuner~\cite{zhang2023mindopt} is a user-friendly, cloud-based platform for tuning arbitrary algorithms through a combination of black-box optimization algorithms.

%Bibliography
\bibliographystyle{unsrt}  
\bibliography{main}  

\section*{Appendix}
\addcontentsline{toc}{chapter}{Appendices}
\appendix

\section{CPLEX 12.09 License Agreement}
A pertinent question that may arise is whether we are authorized to publish benchmark results for CPLEX. As licensed users of CPLEX, we adhere to its Software License Agreement \cite{ibmlicense} including the relevant part:

\begin{quote}
``Licensee may disclose the results of any benchmark test of the Program or its subcomponents to any third party provided that Licensee (A) publicly discloses the complete methodology used in the benchmark test (for example, hardware and software setup, installation procedure and configuration files), (B) performs Licensee's benchmark testing running the Program in its Specified Operating Environment using the latest applicable updates, patches and fixes available for the Program from IBM or third parties that provide IBM products ("Third Parties"), and (C) follows any and all performance tuning and "best practices" guidance available in the Program's documentation and on IBM's support web sites for the Program. If Licensee publishes the results of any benchmark tests for the Program, then notwithstanding anything to the contrary in any agreement between Licensee and IBM or Third Parties, IBM and Third Parties will have the right to publish the results of benchmark tests with respect to Licensee's products provided IBM or Third Parties complies with the requirements of (A), (B) and (C) above in its testing of Licensee's products.
Notwithstanding the foregoing, under no circumstances may Licensee publish the results of benchmark tests run on Oracle Outside In Technology without prior written permission.''
\end{quote}

The agreement permits us to disclose benchmark test results under conditions (A), (B) and (C), which we have diligently met. Specifically, we have shared the configuration used to call CPLEX, ensuring compliance with IBM's guidelines and, in many aspects, exceeding their recommended best practices.

\end{document}